\documentclass[9pt,twocolumn,twoside]{opticajnl}
\journal{opticajournal} 

\setboolean{shortarticle}{true}

\usepackage{macro}
\usepackage{lineno}
\usepackage{xspace}
\usepackage{xcolor}

\title{Ergonomic-Centric Holography: Optimizing Realism, Immersion, and Comfort for Holographic Display}

\author[1,$\dagger$,*]{Liang Shi}
\author[1,$\dagger$]{DongHun Ryu}
\author[1]{Wojciech Matusik}

\affil[1]{Computer Science and Artificial Intelligence Laboratory, Massachusetts Institute of Technology, 32 Vassar St, Cambridge, MA, 02139, USA}
\affil[$\dagger$]{These authors contributed equally to this work.}
\affil[*]{Corresponding author: liangs@mit.edu}

\begin{abstract}
We introduce ergonomic-centric holography, an algorithmic framework that simultaneously optimizes for realistic incoherent defocus, unrestricted pupil movements in the eye box, and high-order diffractions for filtering-free holography. The proposed method outperforms prior algorithms on holographic display prototypes operating in unfiltered and pupil-mimicking modes, offering the potential to enhance next-generation virtual and augmented reality experiences.
\end{abstract}

\setboolean{displaycopyright}{true}

\begin{document}

\maketitle

\section{Introduction}
Computer-generated holography (CGH) creates 3D visuals from 2D wavefront modulation, offering unmatched potential for building accommodation-supporting near-eye displays in thin form factor~\cite{chang2020toward}. Recent progress in machine learning, computational optics, and hardware have substantially improved CGH's image quality, computation speed, and resolution~\cite{Shi2017-np, Maimone2017-sg, Shi2021-iw, Peng2020-an, Chakravarthula2020-jq, Yang2022-yn, Kavakli2023-uu, Zhang2015-ki}, however, ergonomics has yet to receive systematic attention. In particular, we recognize three essential aspects of ergonomics: realism, immersion, and comfort. An ideal CGH shall produce an incoherent out-of-focus response matching how real-world objects defocus, minimize the image quality variation across the theoretical eye box to allow unrestricted pupil movement with motion parallax and reduce sensitivity to eye tracking failure, and simultaneously model high-order diffractions to eliminate optical filtering for designing a slim and comfortable display.

Recent works have tackled each of the aforementioned problems separately. Without modeling high-order diffraction, Choi et al.\cite{Choi2022-lo} and Lee et al.\cite{Lee2022-tz} used temporal time-multiplexing to achieve a natural defocus response. Chakravarthula et al.~\cite{Chakravarthula2022-js} incorporated a dynamic pupil to improve image quality at eccentric pupils in the eye box. Otherwise, Gopakumar et al.~\cite{Gopakumar2021-sc} proposed the high-order gradient descent (HOGD) algorithm to enable optical-filtering-free holographic display for 2D targets. Kim et al.\cite{Kim2022-cd} introduced pupil-HOGD for holographic eyeglasses, adding modeling of a single fixed pupil and support for multi-plane targets under unconstrained defocus responses. Despite their successes, a unified framework that simultaneously addresses the above challenges has not been fully explored.

Here, we propose ergonomic-centric holography (\method), an optimization framework that systematically integrates and advances the merits of previous works to improve the ergonomics of CGH. \method combines layered depth images (LDI)\cite{Shade1998-cs,Shi2022-ih} and incoherent wave propagation\cite{Lee2022-tz} to compute a physically accurate 3D focal stack for supervising hologram optimization. An enhanced HOGD algorithm is developed to support multi-hologram optimization for time multiplexing and dynamic pupil modeling to maintain high image quality over the full eye box.

\begin{figure*}
    \centering
    \includegraphics{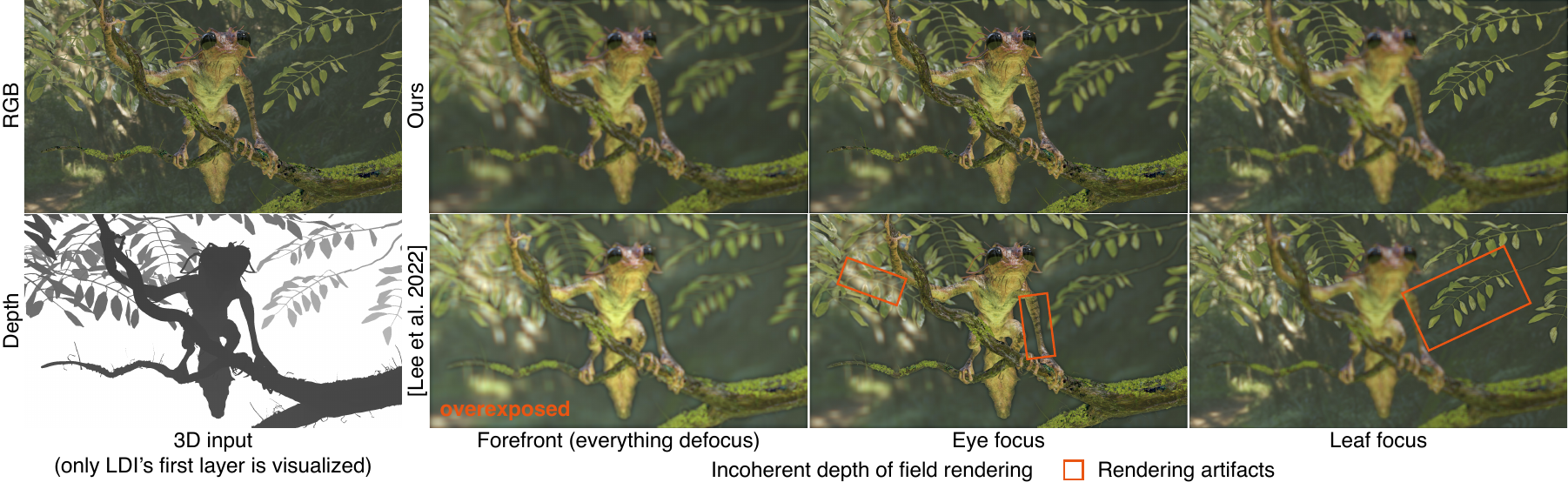}
    \caption{Comparison of the incoherent depth of field rendering. Lee et al.~\cite{Lee2022-tz} apply binary masking and blending on an RGB-D input to process occlusion and accumulate incoherent wavefronts. Their method fails to conserve total energy (the forefront image is noticeably brighter than the eye/leaf focus), produces attenuated background at occlusion boundaries, and allows background light to pass through. The proposed method eliminates these artifacts using ray tracing and LDI representation.}
    \label{fig:incoherent_dof}
\end{figure*}

\method begins by rendering a focal stack that matches real-world defocus response using incoherent wave propagation. Consider a 3D scene with a thickness of $\volumeDepth$ and $\depthLayer$ evenly spaced (for convenience) recording planes (32 in our case), the space-domain incoherent wave propagation kernel $\incohKernelSpace$ for propagating a scene point at depth $\pointDepth$ to the $\aDepthLayer$-th recording plane is given by:
%
\begin{align}
    & \incohKernelSpace(x,y) = \fft\{\cohKernelFreq\}\fft^{*}\{\cohKernelFreq\}\incohKernelMask\label{eq:incoh_kernel_space}\\
    & \cohKernelFreq\left(\fx, \fy\right)= \begin{cases}e^{i \frac{2 \pi}{\wavelen} \sqrt{1-\left(\wavelen \fx\right)^2-\left(\wavelen \fy\right)^2} (\pointDepth-\frac{\aDepthLayer\volumeDepth}{\depthLayer})}, & \text { if } \sqrt{\fx^2+\fy^2}<\frac{1}{\wavelen} \\ 0 & \text { otherwise }\end{cases}
    \label{eq:coh_kernel_freq},
\end{align}
where $\wavelen$ is the wavelength, $\cohKernelFreq$ is the frequency-domain band-limited coherent propagation kernel, and $\incohKernelMask$ is an optional binary mask for space-domain filtering (e.g., conforming the kernel to produce a circular-shaped out-of-focus response, or forcing a deep depth of field). To efficiently and completely model a 3D scene, we use LDI, an advanced multi-layer RGB-depth image representation, to record both foreground and background points (see Supplement 1 for details). For each point, we perform ray tracing with occlusion processing to integrate its incoherent sub-hologram kernel (Eq.~(\ref{eq:incoh_kernel_space})) at each recording plane and render the target focal stack with unquantized per-pixel depth defocus. We show superior image quality over the simple blending and masking method proposed by Lee et al.~\cite{Lee2022-tz} in Fig.~\ref{fig:incoherent_dof}. In practice, we set $\incohKernelMask$ as a circular binary mask to induce a more common circular blur spot (see Supplement 1 Fig. S16 and Video 1 for final rendered examples and focal sweeps).
%

\method enhances the HOGD algorithm with temporal multiplexing and dynamic pupil modeling. Denote the pixel pitch as $\pixel$, the number of orders to model as $\highOrder$, the total number of frames to time multiplex as $\totalFrames$, the total number of circularly-shaped pupils to simultaneously optimize as $\totalPupils$, the radius of a pupil as $\radius$, the support of the eye box as $\{x\in \mathbb {R}: \xmin+\radius \geq x \leq \xmax-\radius; \ y\in \mathbb {R}: \ymin+\radius \geq y \leq \ymax-\radius\}$, where $(x,y)$ is the center of the pupil, and $\xmin, \xmax, \ymin, \ymax$ are the min and max limit along x and y-axis that defines the boundary of the eye box. Throughout optimization, we maintain a set of $\PupilFix$ fixed pupils that forms a uniform pupil sampling grid over the eye box, forcing the energy frequency to be structurally distributed across the whole eye box. At each iteration, we also generate $\PupilRandom = \totalPupils - \PupilFix$ random pupils to account for pupil variations within the lattice of the fixed pupils (see Fig. S7 for a visualization). 

For the $\aFrame$-th SLM pattern $\SLM_{\aFrame}$ and $\aPupil$-th pupil mask $\pupilMask_{\aPupil}$, the field at a distance of $\distToTgtPlane$ is given by:
%
\begin{equation}
\begin{aligned}
& \propagator_{\aPupil}(\SLM_{\aFrame} ; \distToTgtPlane)=\iint \unfiltFreq\left(\fx, \fy ; \SLM_{\aFrame}\right) \unfiltCohKernel_{\aPupil}\left(\fx, \fy ; \distToTgtPlane\right) e^{i 2 \pi\left(\fx x+\fy y\right)} d \fx d \fy, \\
& \unfiltFreq\left(\fx, \fy ; \SLM_{\aFrame} \right)=\sum_{j, k \in \alpha} \fft\left\{e^{i \SLM_{\aFrame}}\right\}\left(\fx+\frac{j}{\pixel}, \fy+\frac{k}{\pixel}\right), \\
& \unfiltCohKernel_{\aPupil}\left(\fx, \fy ; \distToTgtPlane \right)=\cohKernelFreq\left(\fx, \fy ; \distToTgtPlane\right) \sinc\left(\pi \fx \pixel \right) \sinc\left(\pi \fy \pixel \right) \pupilMask_{\aPupil}\\
& \pupilMask_{\aPupil}\left(\fx, \fy; \pupilCenter_{\aPupil_{x}}, \pupilCenter_{\aPupil_{y}}, \radius_\aPupil \right) = \begin{cases} 1, & \text { if } (\fx - \pupilCenter_{\aPupil_{x}})^2 +(\fy - \pupilCenter_{\aPupil_{y}})^2 < \radius_\aPupil^2 \\ 0 & \text { otherwise } \end{cases}
\end{aligned}
\label{eq:image_formation}
\end{equation}
Given a target incoherent focal stack $\{\target_\aDepthLayer| \aDepthLayer=1,\ldots,\depthLayer \}$, we use gradient descent to optimize the batch of time-multiplexed holograms with objective
%
\begin{equation}
    \underset{\{\field_{\aFrame} | \aFrame=1,\ldots,\totalFrames\}}{\text{argmin}} \sum_{\aDepthLayer=1}^{\depthLayer}\sum_{\aPupil=1}^{\totalPupils} \left\lVert \frac{\sqrt{\frac{1}{\totalFrames} \sum_{t=1}^{\totalFrames} \left\lvert \propagator_{\aPupil}(\scaleGlobal\scaleLaser\SLM_{\aFrame} ; \distToTgtPlane_{\aDepthLayer})\right\rvert^2}}{\scalePupil} -  \target_\aDepthLayer \right\rVert,
\label{eq:objective}
\end{equation}
where $\scaleGlobal$ is an optimizable global scale to match the total field intensity with the targets, and $\scaleLaser$ is a non-optimizable per-pixel scale that compensates the non-uniformity of the incident illumination~\cite{Peng2020-an, Shi2021-iw}, $\scalePupil$ is an optional non-optimizable normalization scale that accounts for the pupil size variation (see Supplement 1 for details and improvements we made against previous works).

Our experimental setup uses a Holoeye Pluto SLM with a resolution of 1,080 × 1,920 and 8-bit phase control across visible wavelengths (see Fig. S1 for a schematic rendering). The SLM is mounted on a motorized translation stage (Thorlabs Z825B) to programmably shift position for focus control. Coherent illumination is provided by a FISBA RGBeam fiber-coupled laser with central wavelengths at 632 nm (red), 520 nm (green), and 450 nm (blue). A 4f system with lenses of 80 mm (first) and 200 mm (second) are used to relay and magnify the image to fulfill a full-frame camera sensor (Sony A7III). An optional iris (Thorlabs ID12), mounted on a manual xz-stage (Thorlabs XRN25P/M, XRN-XZ/M), is placed at the Fourier plane to mimic an eye pupil. When the first lens of the 4f system acts as an eyepiece, the central order of the red light diffraction creates an eye box of approximately $6.4\mm \times 6.4\mm$ at the Fourier plane. In unfiltered mode, the iris is absent. In pupil-mimicking mode, the iris is inserted and positioned at different locations. 
%
\begin{figure*}[t]
    \centering    
    \includegraphics[width=\linewidth]{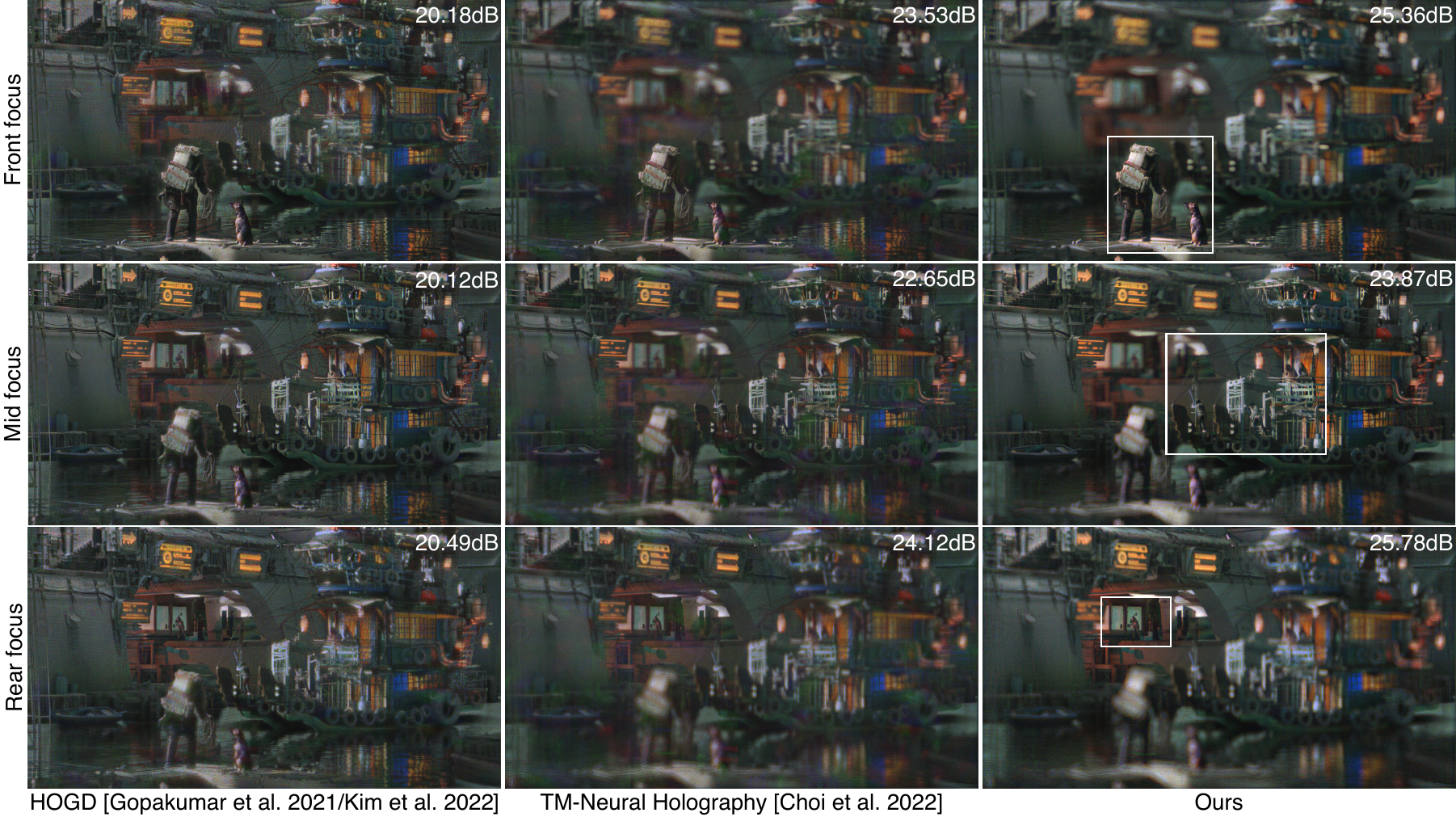}
    \caption{Comparison of 3D CGH algorithms on experimental results captured under unfiltered mode. The camera focuses are marked by white rectangles (plane 4, 16, 28), and the numbers indicate the PSNR with respect to the target image. The captured results are presented in 1200 DPI resolution for close-up examination. Source image: ``PartyTug 6:00AM'' by Ian Hubert.}
    \label{fig:unfiltered_results}
\end{figure*}

During optimization, we consider the central $3 \times 3$ orders ($\highOrder$ = 3). Orders higher than 3 are omitted as they contribute negligibly. For each scene, we optimize for 6 focal planes, typically chosen to have objects of interest in focus. For unfiltered results and pupil-mimicking results, we use 5 and 3 sub-frames for time-multiplexing, respectively (see more details in Supplement 1). 

Figure~\ref{fig:unfiltered_results} compares experimentally captured holograms using EC-H, time-multiplexed neural holography (TM-NH)~\cite{Choi2022-lo}, and HOGD~\cite{Gopakumar2021-sc, Kim2022-cd} in the unfiltered mode. We use our LDI-computed focal stacks to supervise the optimization of TM-NH, as the code to generate their focal stack is not yet publicly available. \method outperforms TM-NH by effectively reducing replicas and rainbow-like artifacts caused by wavelength-dispersed high-order diffractions. This leads to tangibly improved image contrast while preserving the depth-dependent incoherent defocus throughout the 3D volume. Unlike TM-NH, HOGD does not suffer from high-order diffractions. However, it produces coherent defocus responses due to a lack of supervision for out-of-focus regions. For all methods, time multiplexing effectively reduces the speckle noise. Results of additional examples and focal sweep videos can be found in Supplement 1 and Video 1.

\begin{figure*}[t]
    \centering
    \includegraphics[width=\linewidth]{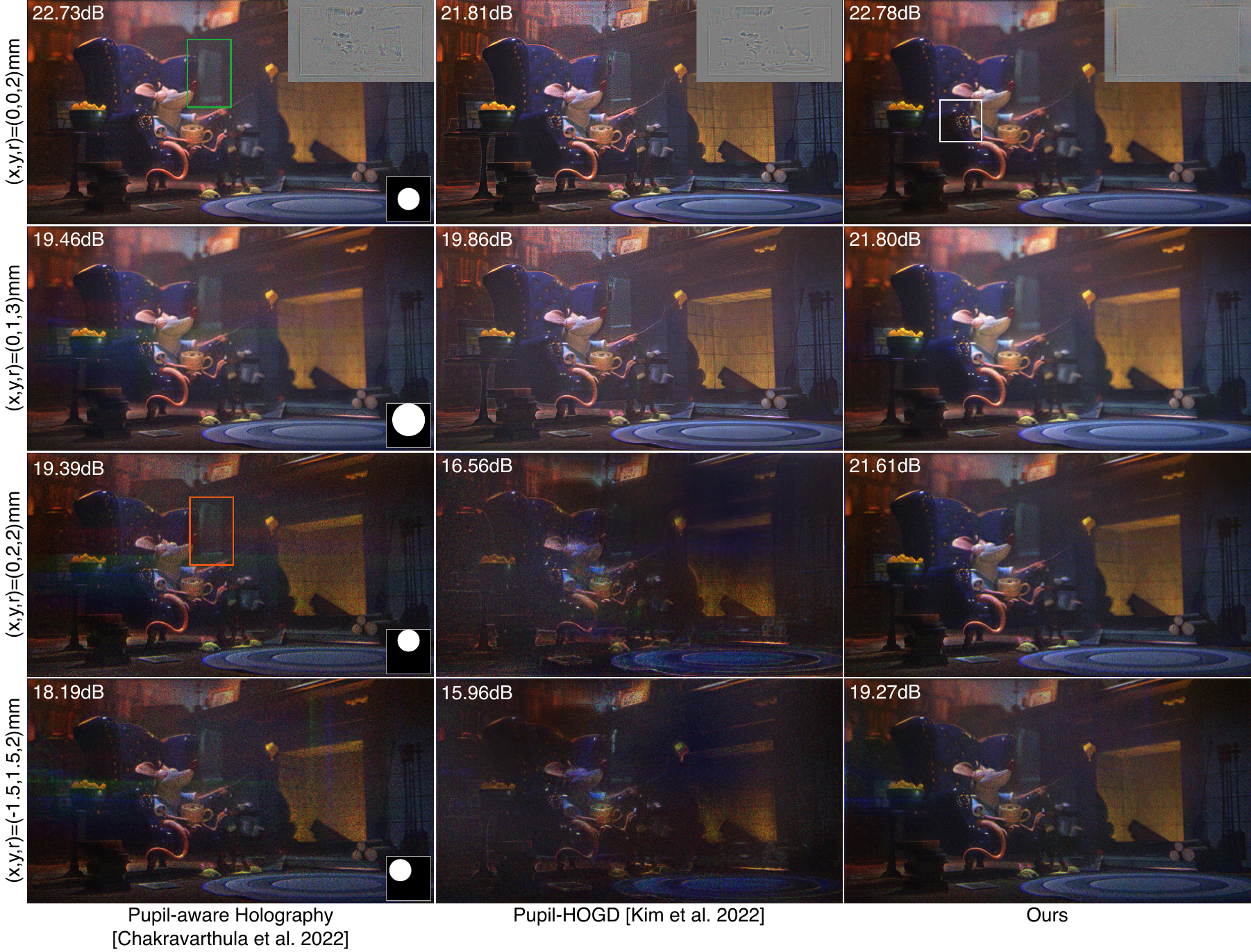}
    \caption{Comparison of 3D CGH algorithms on experimental results captured at various pupil locations and sizes. The bottom right corners in column 1 images mark the pupil position and size used to capture the results in the respective row. The numbers indicate the PSNR with respect to the target image. The captured results are presented in 1200 DPI for close-up examination. Source image: ``Mr. Elephant'' by Glenn Melenhorst. See live capture of the green channel in Supplement Video 2.}
    \label{fig:pupil_results}
\end{figure*}

To optimize for image quality across the eye box, \method consider an $8\times8$ mm eye box given by $(\xmin,\ymin,\xmax,\ymax)$ = $(-4, -4, 4, 4)\mm$, a size bigger than the theoretical maximum as we show modeling high-order diffractions effectively extends the eye box formed by the central diffraction order. We use $\totalPupils=25, \PupilFix=9$ (a $3\times 3$ grid), $\PupilRandom=16$. We set $r=2 \mm$ as the base pupil size to form the uniform sampling grid for the fixed pupils. For the random pupils, their locations are randomly selected within the eye box, and their sizes are scaled between half and twice the base size to account for pupil variations within the fixed pupil lattice. Figure~\ref{fig:pupil_results} compares experimentally captured holograms obtained from \method, pupil-aware holography (PW-H)~\cite{Chakravarthula2022-js}, and Pupil-HOGD~\cite{Kim2022-cd}. Note that the original paper of PW-H optimizes coherent defocus for their two-plane results. As Pupil-HOGD covers reproducing coherent defocus, we upgrade PW-H to reproduce incoherent defocus to emphasize other improvements made by EC-H. 

At eccentric pupils, the Pupil-HOGD method suffers from a significant loss of intensity when the pupil is shifted and reduced to the extent when it fails to fully encompass the DC term. This is evident in the transition from a 6mm pupil in row 2 to a 4mm pupil in row 3, both shifted to the center top of the eye box. This loss occurs as Pupil-HOGD solely regularizes image quality at the center pupil, causing an imbalanced energy distribution in the frequency domain (see Supplement 1). Alternatively, PA-H exhibits pronounced rainbow-like artifacts with reduced contrast, a faster decay in image brightness (see row 3), and a stronger reduction in the extent of reproduced defocus blur compared to EC-H (see the orange box in column 1, row 3 versus the green box above, with the same regions in column 3). They are caused by PA-H's vulnerability to high-order diffractions and the absence of fixed pupils during optimization, which further push the energy spectrum structurally to the mid/high frequencies (see row 1 bottom right for holograms and Supplement 1 for spectrum analysis). \method better maintains the image intensity and quality as the pupil moves away and reduces. It also produces artifact-free images outside the central-order-created eye box. The extended eye allows the perception of noticeable motion parallax (see Supplement 1/Video 3). Additional results can also be found in Supplement 1.

In conclusion, we demonstrate \method can effectively improve the display ergonomics for computer-generated holography via synergizing and advancing efforts made in recent works (see discussions and limitations in Supplement 1).
%
Future works can be built on top of \method to further improve its performance. First, the space-bandwidth product (i.e., etendue) that determines the product of the eye box and field-of-view shall be further enhanced for more immersive VR/AR experiences. Recent applications of high-resolution random ~\cite{Kuo2020-Et, park2019-Si, yu2017ultrahigh} or engineered ~\cite{Baek2021-Nr} phase masks for etendue expansion can be incorporated for joint optimization. Second, \method can be accelerated using deep neural networks for real-time hologram generation~\cite{Shi2021-iw, choi2021-nr, Yang2022-yn}. Third, \method can be extended to model multi-color holograms~\cite{Kavakli2023-kv} to support modulation of poly-chromatic illumination for higher image brightness without using more powerful lasers. 

\begin{backmatter}

\bmsection{Acknowledgments} We thank Byounghyo Lee for sharing their incoherent focal stack rendering code for comparison. L.S. is supported by Meta Research PhD Fellowship; D.R. is supported by MIT EECS Alumni Fellowship. 

\bmsection{Disclosures} The authors declare no conflicts of interest. 

\bmsection{Data Availability} Source code and data needed to evaluate the conclusions will be made timely and publicly available at: https://github.com/liangs111/ergonomic-centric-holography

\bmsection{Supplemental document} See Supplement 1 for supporting content.

\end{backmatter}

\bibliography{sample}

\bibliographyfullrefs{sample}

\ifthenelse{\equal{\journalref}{aop}}{%
\section*{Author Biographies}
\begingroup
\setlength\intextsep{0pt}
\begin{minipage}[t][6.3cm][t]{1.0\textwidth} 
  \begin{wrapfigure}{L}{0.25\textwidth}
    \includegraphics[width=0.25\textwidth]{john_smith.eps}
  \end{wrapfigure}
  \noindent
  {\bfseries John Smith} received his BSc (Mathematics) in 2000 from The University of Maryland. His research interests include lasers and optics.
\end{minipage}
\begin{minipage}{1.0\textwidth}
  \begin{wrapfigure}{L}{0.25\textwidth}
    \includegraphics[width=0.25\textwidth]{alice_smith.eps}
  \end{wrapfigure}
  \noindent
  {\bfseries Alice Smith} also received her BSc (Mathematics) in 2000 from The University of Maryland. Her research interests also include lasers and optics.
\end{minipage}
\endgroup
}{}

\end{document}